\documentclass{moriond}

\def\be{\begin{equation}}
\def\ee{\end{equation}}
\def\bea{\begin{eqnarray}}
\def\eea{\end{eqnarray}}

\newcommand{\Photo}{\includegraphics[height=35mm]{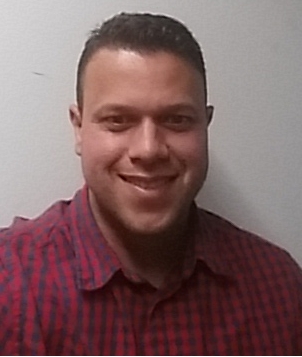}}

\begin{document}
\vspace*{4cm}
\title{Dark Matter Overview: Collider, Direct and Indirect Detection Searches}

\author{Farinaldo S. Queiroz }

\address{Max-Planck-Institut f\"ur Kernphysik, Saupfercheckweg 1, 69117 Heidelberg, Germany}

\maketitle\abstracts{
The complementarity of direct, indirect and collider searches for dark matter has improved our understanding concerning the properties of the dark matter particle. I will review the basic concepts that these methods rely upon and highlight what are the most important information they provide when it comes down to interpret the results in terms of Weakly Interacting Massive Particles (WIMPs). Later, I go over some of the latest results emphasizing the implications to dark matter theory in a broad sense and point out recent developments and prospects in the field. }

\section{Introduction}
\label{sec:1}

\begin{figure}[!h]
\centering
\includegraphics[scale=0.5]{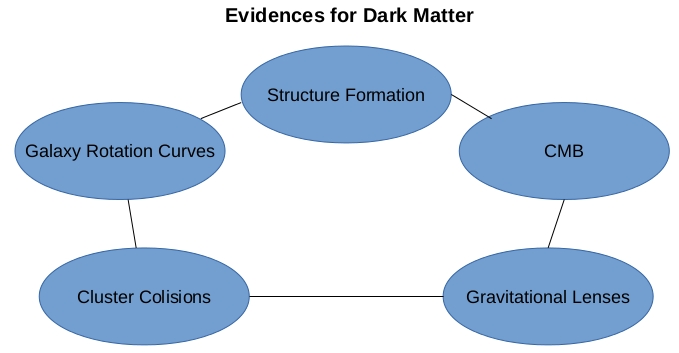}
\caption{Collection of the five most important evidences for dark matter: structure formation, galaxy rotation curves, cluster collisions, Cosmic Microwave Background (CMB) and Gravitational lenses.}
\label{evidence}
\end{figure}

It is well known that dark matter accounts for about 85\% of the matter content of the universe and roughly 27\% of the entire energy density. Moreover, it is common knowledge that dark matter played an important role in the expansion history of the universe, specially in the formation of structures we observe today, such as galaxies and cluster of galaxies. The presence of dark matter has been ascertained through its gravitation effects by several observations as shown in Fig.\ref{evidence}. Unfortunately we dispose of no solid evidence for dark matter based on its interaction with ordinary matter that collider, direct detection and indirect detection methods reply upon. Thus, the puzzling question is: What is the nature of dark matter? We know dark matter is out there, but what kind of particle is dark matter made of? In more specific terms, we would like to know at some point the spin, mass and quantify the interaction strength of the dark matter particles with the standard model ones if any. Those are all open questions which might take a very long time to be answered even if a robust dark matter signal is observed today \cite{Roszkowski:2016bhs}. As Max Planck (1858-1947) once said ``An experiment is a question which science poses to Nature, and a measurement is the recording of Nature's
answer'', and indeed we have recorded many important answers along the past decades that helped us rule out what dark matter particles mostly likely cannot be, as well as identify some properties dark matter particles could have that would yield a signal within reach of current and planned experiments, while simultaneously fitting the observations.

A variety of observations going from structure formation to Cosmic Microwave Background data helped us infer some properties of the dark matter particles: (i) Structure formation tell us that dark matter particles could not have had a large free-streaming during the period of structure formation which took place around $10^{12}$~sec or so \cite{Cornell:2013rza}. Hence if dark matter particles belonged to a thermal history throughout the universe expansion, they cannot be very light ($\ll 1\, keV$); (ii) additionally, searches for electrically charged stable particles have occurred with null results, which resulted into stringent limits on models often called charged or milicharged dark matter \cite{SanchezSalcedo:2010ev}, which lead us to believe that dark matter is effectively electrically neutral; (iii) Among other observations, the precise measurement of the power spectrum of the cosmic microwave background radiation infers that the cold dark matter component of the universe should account for 27\% of the energy budget of the universe \cite{Ade:2015xua}. This is very important because large regions of parameter space from a multitude of models have been ruled out, since they predict overabundant dark matter; (iv)  We do observe gravitational effects of the dark matter in our universe today, so dark matter particles should be stable at cosmological scales. It means that their lifetime should be much larger than the age of the universe ($\sim 4 \times 10^{17}$ sec). Though much stronger constraints can be derived using gamma-ray, neutrino and cosmic-ray data \cite{Audren:2014bca,Queiroz:2014yna,Baring:2015sza,Mambrini:2015sia,Giesen:2015ufa}; (v) Cluster collisions indicate that dark matter particles are not strongly interacting particles \cite{Clowe:2006eq}. Actually the constraint from these collisions are rather loose leaving room for strongly interacting scenarios \cite{Hochberg:2014dra,Hansen:2015yaa}. 

Anyways, if one could wish for a dark matter candidate easy to be incorporated in particle physics models, and able to address the five Nature's answers above, plus predicting signals within sensitivity of current or future experiments, which candidate would that be? The answer is WIMPs (Weakly Interacting Massive Particles). I would not dare to say they are the best dark matter candidates, but they are without shadow of a doubt the most popular, and are the focus of this brief review. The question in order is, how do we search for these particles? The three main methods to search for WIMPs are known as: collider, direct and indirect detection which we will review below and highglight some recent developments.
\section{Collider Searches}

\begin{figure}[!h]
\centering
\includegraphics[scale=0.5]{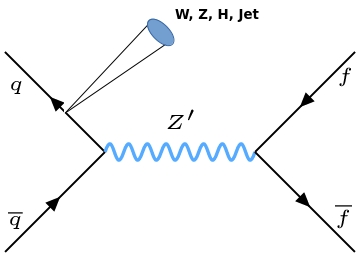}
\caption{Mono-X searches for dark matter exhibited for the case of s-channel vector mediators.}
\label{monoX}
\end{figure}

Since dark matter particles are assumed to be electrically neutral and cosmologically stable, at colliders they are simply refereed as missing energy. Some sort of heavy neutrinos. Hence, collider searches for dark matter are based on the detection of the visible counterpart of the signal, such as jets and charged leptons. An important aspect regarding collider searches for dark matter that needs to be stressed is the fact that colliders generally speaking offer a complementary and important probe for dark matter, but they cannot determine if what they see is the dark matter of the universe, since any neutral particle that decays outside the detector can be seen as missing energy. Only direct and indirect detection methods provide a way to confirm whether a potential signal is truly due to dark matter. Fig.\ref{monoX} illustrates what is often referred as mono-X searches, which accounts for mono-Z,mono-H, mono-jet searches for dark matter. They all have their virtues and setbacks when comes down to probing the parameter space of dark matter models.

The common approach in this endeavour is to use effective operators to describe the interaction between the dark matter particle and fermions. In the case of Dirac and Majorana fermions mediated by a vector boson, as shown in Fig.\ref{monoX}, the relevant operators are,\\

Dirac Fermion: $\frac{1}{\Lambda^2} \bar{\chi}\gamma_{\mu}\chi \bar{q}\gamma_{\mu} q$ + $\frac{1}{\Lambda^2} \bar{\chi}\gamma_{\mu}\gamma_5\chi \bar{q}\gamma_{\mu}\gamma_5 q$,\\

Majorana Fermion: $\frac{1}{\Lambda^2} \bar{\chi}\gamma_{\mu}\gamma_5\chi \bar{q}\gamma_{\mu}\gamma_5 q$.\\

The use of effective operators makes the interpretations among collider, direct and indirect detection observables fairly simpler than in the context of simplified dark matter models or UV complete theories. However, its simplicity comes at a price, which is the loss of resonance effects and overestimated limits in the regime which the momentum transfer is larger than the mediator mass. Let me explain better. Ignoring the width in the propagator for now, in the case of vector currents, the simplified lagrangian $L \supset [g_{\chi}\bar{\chi}\gamma_{\mu}\chi + g_q \bar{q}\gamma_{\mu} q]Z^{\prime}$ in the limit $M_{Z^{\prime}} \gg Q$, where Q is the momentum transfer, results into a signal amplitude proportional to,

\begin{equation}
\frac{g_q g_{\chi}}{M_{Z^{\prime}}^2 -Q ^2} \sim \frac{g_q g_{\chi}}{M_{Z^{\prime}}^2}\left(1+ \frac{Q^2}{M_{Z^{\prime}}^2}\right).
\end{equation}

That said one can differentiate three regimes:\\

(i) {\it Effective field theory is valid}\\

In the effective field theory approach one can match the unknown scale to the mediator mass,  with $\Lambda=\frac{M_{Z^{\prime}}}{\sqrt{g_q g_{\chi}}}$. So that $\Lambda$ encompasses the couplings and the mediator mass as well as the quantum number of $\chi$ under the new gauge group which $Z^{\prime}$ is originated from. This matching is valid when $M_{Z^{\prime}} \gg Q$. For the LHC at 13TeV,  it means that the effective theory approach is robust for mediators as heavy as 10TeV \cite{Abercrombie:2015wmb}.\\

(ii) {\it Effective field theory underestimates observables}\\
 
When $M_{Z^{\prime}} \sim Q$, the $Z^{\prime}$ is produced on shell, leading to large production rates. In this regime the use of effective theory yields underestimated limits. Note that in many particle physics models, mainly those related to vector mediators, the right relic abundance is obtained through resonance effects. Therefore, the effective field theory approach actually fails twice.  \\

(iii) {\it Effective field theory breaks down}\\

When $Q > M_{Z^{\prime}}$, the effective field theory approach breaks down, overestimating the signals and bounds.\\

Thus, one has to carefully interpret LHC data to constrain dark matter models through effective operators. To avoid misuse, missing transverse energy searches from the LHC are planned to be interpreted in terms of simplified dark matter models. See \cite{Boveia:2016mrp} for more details. Keep in mind that, in a given particle physics model, specially those that rely on s-channel production mechanisms, mono-X searches often do not provide the most efficient way to probe dark matter models. Instead, dijet and dilepton resonance searches give rise to the most restrictive bounds. For recent and extensive discussions on the topic see \cite{DeSimone:2016fbz,Englert:2016joy,Alves:2015mua}. I will now discuss dark matter searches that are subject to larger uncertainties, namely direct and indirect detection.

\section{Direct Detection}

\begin{figure}[!h]
\centering
\includegraphics[scale=0.25]{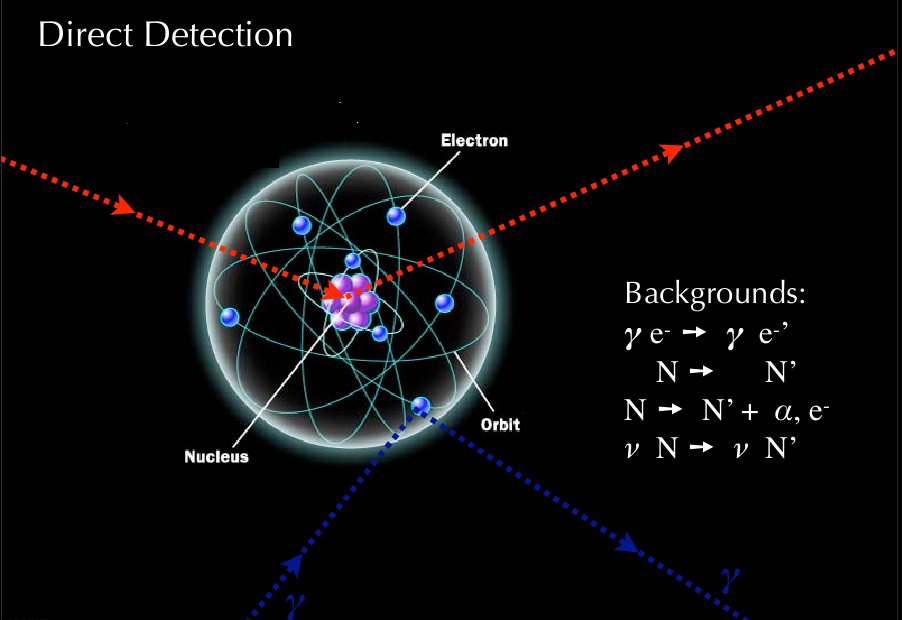}
\includegraphics[scale=0.55]{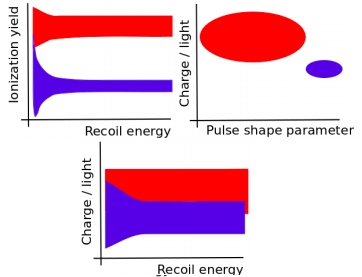}
\caption{{\it Left}: Illustrative dark matter-nucleus scattering which direct detection experiments are based on. {\it Right:} Possible signal-background discriminating variables  used in Germanium, liquid XENON and liquid ARGON detectors.}
\label{DD}
\end{figure}

Since the presence of dark matter in our galaxy is inferred through its gravitational effects by a multitude of observations, direct detection experiments hope to observe dark matter scattering off nuclei targets, which are placed in underground laboratories to shield the detector from cosmic-rays induced events. These searches for dark matter are based on measuring the energy deposited by a dark matter particle in the scattering process with nuclei as it is schematically shown in Fig.\ref{DD}. The dark matter- nuclei scattering rate can be written as,

\begin{equation}
\overbrace{\frac{dR}{dE} (E,t)}^{Scatt.\, Rate} =  {\color{yellow} \underbrace{N_T }_{Target\, Dependence}} {\color{green} \overbrace{\frac{\rho_{\chi}}{m_{\chi}}}^{Number\, density}} {\color{red} \int_{v_{min}}}{\color{blue} \underbrace{\frac{d\sigma }{dE}(v,E)}_{Diff.\,\, Cross\, Section}}\,{\color{red} v \overbrace{f_E(\overrightarrow{v},t)}^{veloc.\, distribution} d^3\overrightarrow{v}},
\end{equation}where $N_T$ is the number of target nuclei per kilogram of the detector, $\rho_{\chi}$ the local dark matter density ($\rho_{\chi} = 0.3\, GeV/cm^3$), $m_{\chi}$ the dark matter mass, $\overrightarrow{v}$ the velocity of the dark matter particle relative to the Earth, $f_E(\overrightarrow{v},t)$ velocity distribution of the dark matter in the frame of the Earth, i.e. the probability of finding a dark matter particle with velocity $v$ at a time $t$, and $v_{min}=\sqrt{m_N E/(2\mu^2)}$ is the minimum dark matter speed which can cause a recoil of energy detectable by a given experiment, with $\mu = m_{\chi} m_N/ (m_{\chi} + m_N)$ being the dark matter-nucleus reduced mass ($m_N$ is the nucleus mass), $d\sigma/dE(v,E)$ the differential cross-section for the dark matter-nucleus scattering as follows,

\begin{equation}
\frac{d\sigma}{dE} = \frac{m_N}{2\mu^2 v^2} (\sigma_{SI} F^2(q) + \sigma_{SD} S(q)),
\end{equation}with $F^2(q),S(q)$ being the spin-independent and spin-dependent form factors respectively listed in \cite{Duda:2006uk}.

The key aspect of direct detection searches for dark matter is not the measurement of the recoil energy itself, but how this energy is distributed. In Fig.\ref{DD} ( adapted from \cite{Undagoitia:2015gya}), I show how one could separate signal from background using ionization yield as used in Germanium detectors, e.g. SuperCDMS \cite{Agnese:2015nto}, charge/light x recoil energy collection as done in experiments such as XENON1T and LUX which use liquid XENON \cite{Aprile:2015uzo,Akerib:2015rjg} and charge/light ratio x pulse shape as performed in liquid Argon detectors as Darkside \cite{Agnes:2015ftt}. Hence, using discriminating variables one can disentangle signal from background and concretely search for dark matter scatterings. Be aware that there are important basic concepts (and/or assumptions) built-in those searches, namely:

\begin{itemize}
\item There is a smooth halo of dark matter particles in our galaxy described by a Maxwell Velocity distribution.
\item Due to the rotation of the Galactic Disk the solar system experiences an effective WIMP
Wind, which leads to an annual Modulation due to Earth's orbital motion.
\item The nucleus is treated as a hard sphere described by the Helm form factor \cite{Duda:2006uk}.
\item The scattering is elastic.

\end{itemize}

There are multiple studies where the impact of different velocity distributions , form factors and ineslatic scatterings are analysed. However, both cosmological simulations including baryons and lattice QCD  studies seem to tell us that the dark matter-scattering process is well described by a Maxwellian velocity and Helm form factor \cite{Sloane:2016kyi,Kelso:2016qqj,Anand:2013yka}. The elasticity of the scattering has something do with the particle physics model in the case where excited dark matter states exist \cite{TuckerSmith:2001hy} though.

In summary, if a signal (e.g. annual modulation and/or excess of nuclear recoil events) is observed, we can related the scattering cross section and mass of the dark matter particle to its local density. For this reason direct detection can truly discover the dark matter particle that permeates our galaxy.

\section{Indirect Detection}

Dark matter particles that populate our universe in galactic and extragalactic scales may self-annihilate and produce a flux of gamma-rays, cosmic-rays, neutrinos, anti-matter which can appear as an excess over the expected background. The flux originated from dark matter annihilation should be proportional to the number density squared of particles, i.e. $\rho_{\chi}^2/m_{\chi}^2$, to the annihilation cross section $\sigma v$, to the element of volume of the sky observed accounted by $\Omega$, and the number of particles of interest produced per annihilation ($dN/dE$). Hence, it can we written as,

\begin{equation}
\overbrace{\frac{d\Phi}{d\Omega dE}}^{Diff. Flux} = {\color{blue} \frac{ \overbrace{ \sigma v }^{Anni.\, Cross\, Section}}{8\pi m_{\chi}^2}} \times {\color{green} \underbrace{\frac{dN}{dE}}_{Energy\, Spectrum}} \times {\color{red} \int_{l.o.s} ds} {\color{red}  \underbrace{\rho^2 (\overrightarrow{r}(s,\Omega))}_{Dark\, Matter\, Distribution}},
\label{eq:flux}
\end{equation}where $\Omega$ is truly the solid angle of the region of interest, $dN/dE$ is the energy spectrum (e.g. the number of photons produced per annihilation in case of gamma-rays), and $\rho (\overrightarrow{r}(s,\Omega))$ is the dark matter density which should integrated over the line of sight (l.o.s) from the observer to the source, which is often assumed to be described by either a Navarro-Frenk-White,
\begin{equation}
\rho(r) \propto \frac{r_s}{r[1+ r/r_s]^2},
\end{equation}or Einasto profile,
\begin{equation}
\rho(r) \propto exp \left[ \frac{-2.0}{\alpha} \left(\,  (r/r_s)^{\alpha}-1 \right) \right],
\end{equation}where $r_s=20$~kpc is the scale radius of the halo, and $\alpha=0.17$.

From Eq.\ref{eq:flux} we see that indirect detection is sensitive to the dark matter density distribution, annihilation cross section and mass. These are complementary information to collider and direct detection searches. For example, if a signal is seen in direct detection and the mass and scattering cross section are inferred with a certain precision we can use this information to determine the dark matter density profile through indirect detection. Note that, the task to pinpoint the dark matter quantum numbers is much more difficult, since in a particle physics model, the direct and indirect detection observables are not necessarily strongly correlated. Even after including some collider input, the nature of dark matter may remain unknown.  Only in cases where there are strong correlations between the parameters that set the collider, direct and indirect detection observables the nature of dark matter particle can be unveiled.

Anyhow, indirect search for dark matter has evolved tremendously due to the amount of data at our disposal. Today, we have a much better understanding of cosmic-ray propagation \cite{Profumo:2013yn}, better handle on the energy spectrum with the inclusion of electroweak and QCD corrections \cite{Ciafaloni:2010ti,Bringmann:2012vr,Bringmann:2015cpa}, and the catalog of gamma-ray sources has vastly enlarged, for instance, and we hope that the recent improvements in the three aforementioned methods will help us unmask the nature of dark matter. After reviewing basic aspects of indirect dark matter detection we will briefly discuss recent signals which have been attributed to dark matter annihilations or decays. 

\subsection{Gamma-ray Excesses}

\begin{figure}[!h]
\centering
\includegraphics[scale=0.6]{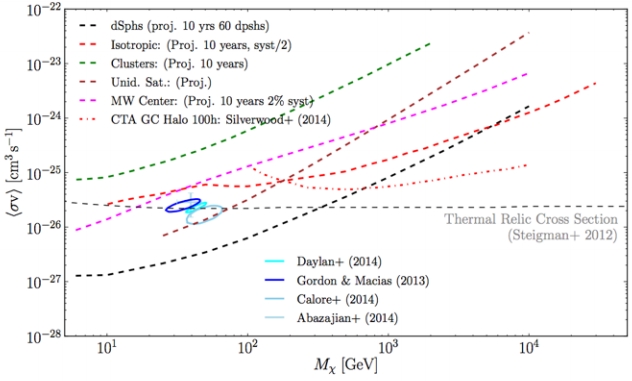}
\label{dwarfproj}
\caption{10 Years limit projections from the observation of dwarf galaxies using Fermi-LAT. Taken from \url{http://fermi.gsfc.nasa.gov/science/mtgs/symposia/2015/program/monday/session1/JRacusin.pdf}}
\end{figure}

Due to the dim signal expected from dark matter annihilation and sizeable uncertainties in the astrophysical background, indirect detection searches for dark matter give rise to a multitude of excesses, which later led to a better understanding of the associated background. Nowadays, we have two excesses in the gamma-ray band, which have been attributed to dark matter annihilation. One observed in the galactic center \cite{Hooper:2010mq,TheFermi-LAT:2015kwa} and other in the recently discovered dwarf galaxy known as Reticulum II \cite{Geringer-Sameth:2015lua}. Both excesses can be explained by the same dark matter particle annihilating into $\bar{b}b$ quarks with $\sigma v \sim 10^{-26} cm^3s^{-1}$, $m_{\chi} \sim 30$GeV. Which is an intriguing coincidence. Although, due to the large uncertainties concerning the dark matter content in this dwarf galaxy is unclear whether the Reticulum II anomaly is actually supporting the galactic center excess \cite{Drlica-Wagner:2015xua}. Moreover, the dark matter properties that can accommodate these excesses should also produce an a gamma-ray signal in other dwarf galaxies, for instance. Current results from Fermi-LAT collaboration already rules out most of the region in the annihilation cross section {\it vs} mass plane that can accommodate the excesses \cite{Ackermann:2015zua}. With the discovery of nearby dwarf galaxies and better statistics, dwarf galaxies are the most promising method to test whether those signals are actually arising as a result of dark matter annihilation. In the foreseeable future we should start seeing excesses in dwarf galaxies, otherwise the upcoming limits based on a stack of dwarf galaxies will place severe limits on the dark matter interpretation of the galactic center excess as shown in Fig.\ref{dwarfproj}. From a more optimistic view though, we might actually be on the verge of a dark matter discovery.

\subsection{keV Line Emission}

A $3.5$~keV line emission has been observed in both stack of 73 clusters of galaxies \cite{Bulbul:2014sua}, galactic center \cite{Jeltema:2014qfa}, in the M31 galaxy and the Perseus galaxy cluster using using data from the XMM-Newton satellite \cite{Boyarsky:2014jta}. The galaxy cluster teams argue that we there should be no atomic
transitions in thermal plasma at this energy, thus such x-ray emission should have an exotic origin such as from dark matter decay with a lifetime of $6-8\times 10^{25}$~sec and mass of $7$~keV, as naturally predicted in sterile neutrino models \cite{Dodelson:1993je,Abazajian:2014gza}. However, \cite{Jeltema:2014qfa} disputed these observations with the suggestive title ``dark matter searches going bananas'' in reference to a possible unaccounted potassium x-ray emission that could absorb the signal. Instead of getting into the debate driven by the authors, let me comment on the facts. In \cite{Jeltema:2014qfa}, the authors focused on the 3-4keV energy range and adopted a public version of the tool used to model the line emission, differently from the previous papers. 

As similarly occurs at colliders, when an experiment such ATLAS observes a strong signal, e.g. the diphoton excess, the event should be checked (seen) by CMS and vice-versa. In the context of indirect dark matter, what we do is to look at a different target, since the same dark matter producing the $3.5$~keV line in clusters should be present in dwarf galaxies for instance. That is exactly what the groups have done later on. They both looked at the Draco dwarf galaxy since it is a classical target for dark matter searches. In \cite{Ruchayskiy:2015onc} they reported null result from Draco, however they affirm that to be consistent with
the 3.5 keV line observed in the stack of galaxy clusters at 95\% C.L. In \cite{Jeltema:2015mee}, on the order hand, the authors concluded that the non-observation of a x-ray line in Draco actually excludes the 3.5 keV line at 99\% C.L. Moreover, in \cite{Phillips:2015wla} the authors re-analysed the x-ray emission lines claimed in \cite{Jeltema:2014qfa}, and concluded that in fact the 3.5~keV feature can indeed be absorbed and there is no need to invoke new physics effects. Unfortunately, there is not enough x-ray data from other dwarfs so that one could perform a stacked analysis to bring the hammer down on this debate. Future observations of Astro-H are expected to clarify the nature of the x-ray line emission. 


After discussing debatable indirect detection signals from dark matter, I will focus now on the observation of dwarf galaxies using Fermi-LAT as a method to probe gamma-ray and neutrino lines from dark matter annihilation.

\subsection{Gamma-ray Limits on Neutrino Lines From Dark Matter Annihilation}

\begin{figure}[!h]
\centering
\includegraphics[scale=0.25]{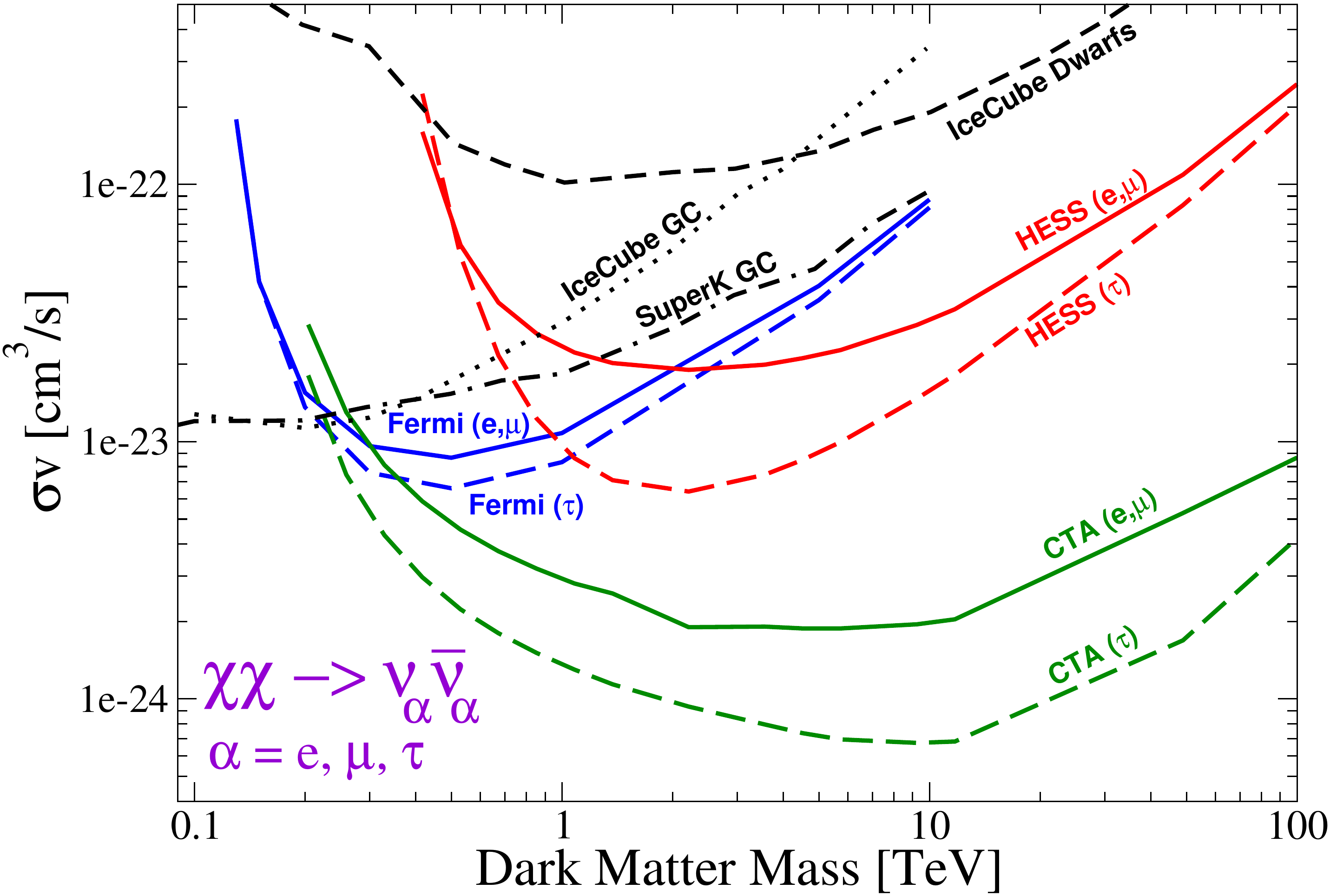}
\includegraphics[scale=0.5]{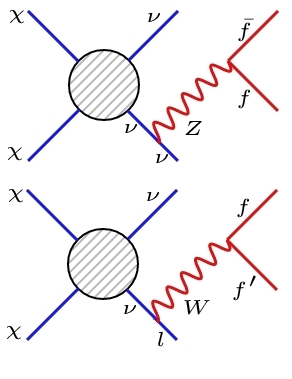}
\label{neutrinoline}
\caption{Dark matter annihilation purely into neutrinos also gives rise to a continuous gamma-ray emission which can be probed with Fermi-LAT, H.E.S.S. and CTA. Notice that in principle, one could distinguish the neutrino flavors using gamma-rays.}
\end{figure}

Dark matter particles might self-annihilate into standard model particles such quarks, charged leptons and neutrinos. Quarks and charged leptons produce a significant amount of continuous gamma-ray emission after final state radiation and hadronization processes are accounted for, which does not occur for final state neutrinos. Thus, if you have a pair of dark matter particles, in the WIMP mass regime, annihilating purely into neutrinos, which detector would you use to search for this dark matter particle? Probably the first thing that comes to mind is IceCube, Super-K etc. Nevertheless, monochromatic neutrinos from dark matter annihilations are accompanied by a gamma-ray spectrum generated by electroweak corrections. Thus we can use gamma-ray telescopes to probe this dark matter particle. It turns out as displayed in Fig.5 that gamma-rays indeed, for masses above $200$~GeV, result into the strongest limits \cite{Queiroz:2016zwd}. It is known that final state tau leptons produce a harder gamma-ray spectrum than electrons. Therefore, the electron-neutrino and tau-neutrino final states yield different gamma-ray spectra, since a electron-neutrino is converted to an electron via W exchange, whereas a tau-neutrino is converted to a tau lepton via W exchange (right side of Fig.5). In other words, if we ever reach this level of precision to discriminate the difference between the $\nu_{e,\mu}$ the $\nu_{\tau}$ curves, gamma-rays offer a promising avenue to distinguish final state neutrino flavors, which is something not possible at neutrino detectors. Note that this result does not undervalue the role of neutrino telescopes as far as detecting dark matter annihilations into neutrinos is concerned, since only them are capable of determining whether the signal is truly a neutrino line.  What this result is showing us is that if a neutrino line is observed in Icecube/Super-K in the mass range of interest, gamma-ray telescopes should also see the corresponding gamma-ray counterpart. There are caveats though, for instance if the annihilation on shell species which then decay into neutrinos, softening the gamma-ray yield. Anyway, leaving such particular cases aside, Fig.\ref{neutrinoline} shows that have already entered into a new era where gamma-ray detectors are actually more sensitive than neutrino detectors to neutrino lines from dark matter annihilation.

Another interesting outcome of the inclusion of electroweak corrections has to do with gamma-ray line searches as I discuss below.

\subsection{Extending Fermi-LAT and H.E.S.S. Limits on Gamma-ray Lines}

Knowing that the Fermi-LAT energy limit is 500GeV. Is Fermi-LAT sensitive to a $2$~TeV dark matter particle annihilating into two photons? At first sight no. Although, gamma-rays may also radiate $W^{\pm}$ gauge bosons which then decay and generate gamma-rays at lower energies, below $500$~GeV, i.e. within Fermi-LAT sensitivity. Thus one can derive new limits, though not as restrictive as those coming from spectral line analysis, on a mass region previously not probed by Fermi-LAT in the context of gamma-ray line searches. The same idea can be applied to H.E.S.S. instrument, whose energy limit extends up to $20$~TeV. In Fig.6, along with the unitary bound taken from \cite{Beacom:2006tt}, I show that both Fermi-LAT and H.E.S.S. limits can be extended to masses much larger than their energy limit using gamma-ray observations of dwarf galaxies for Fermi-LAT, and galactic center for H.E.S.S.. In particular, these are the first limits on gamma-ray lines from dark matter
annihilation for masses above 20TeV. For more details see \cite{Profumo:2016idl}.

\begin{figure}[!h]
\centering
\includegraphics[scale=0.25]{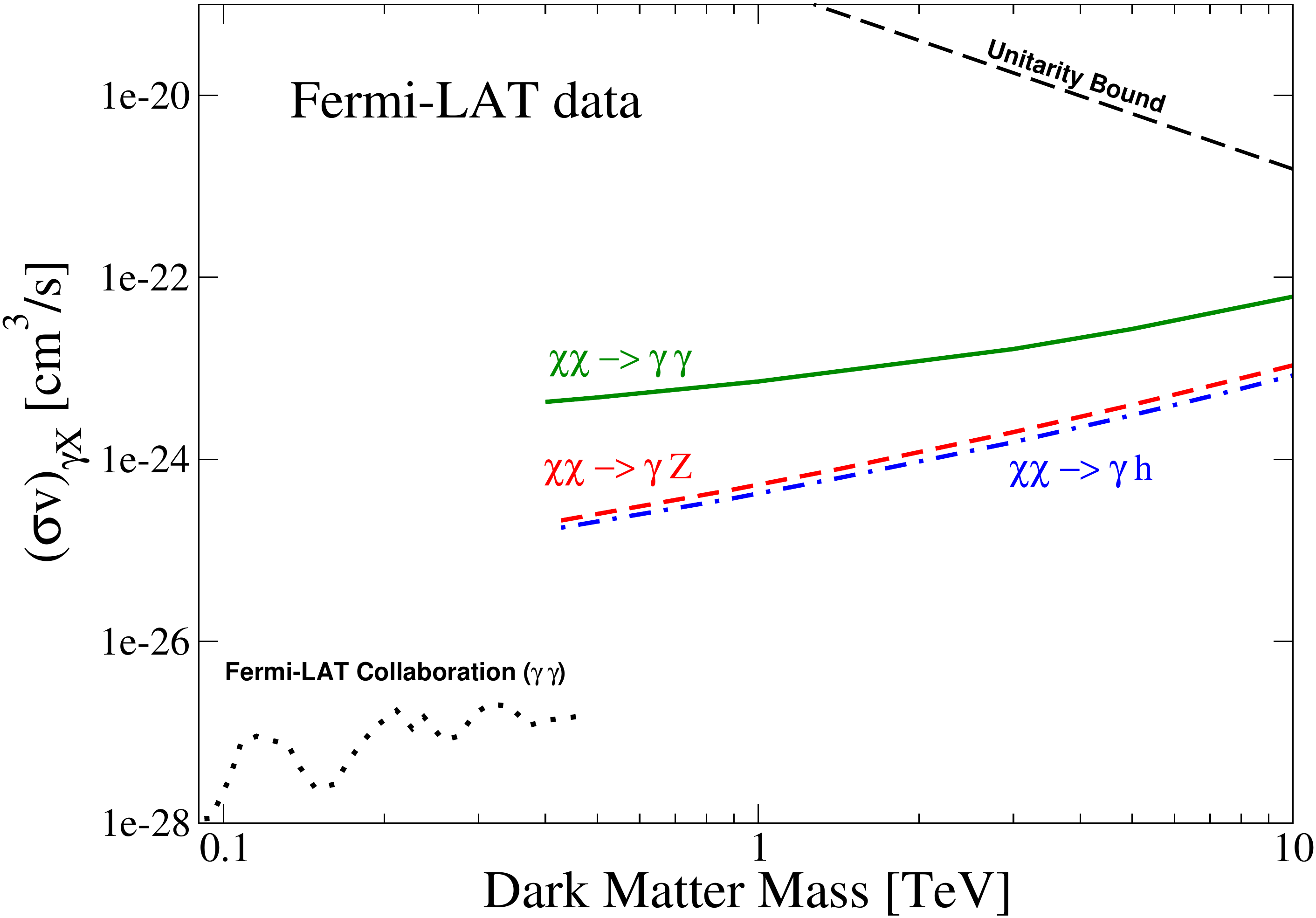}
\includegraphics[scale=0.25]{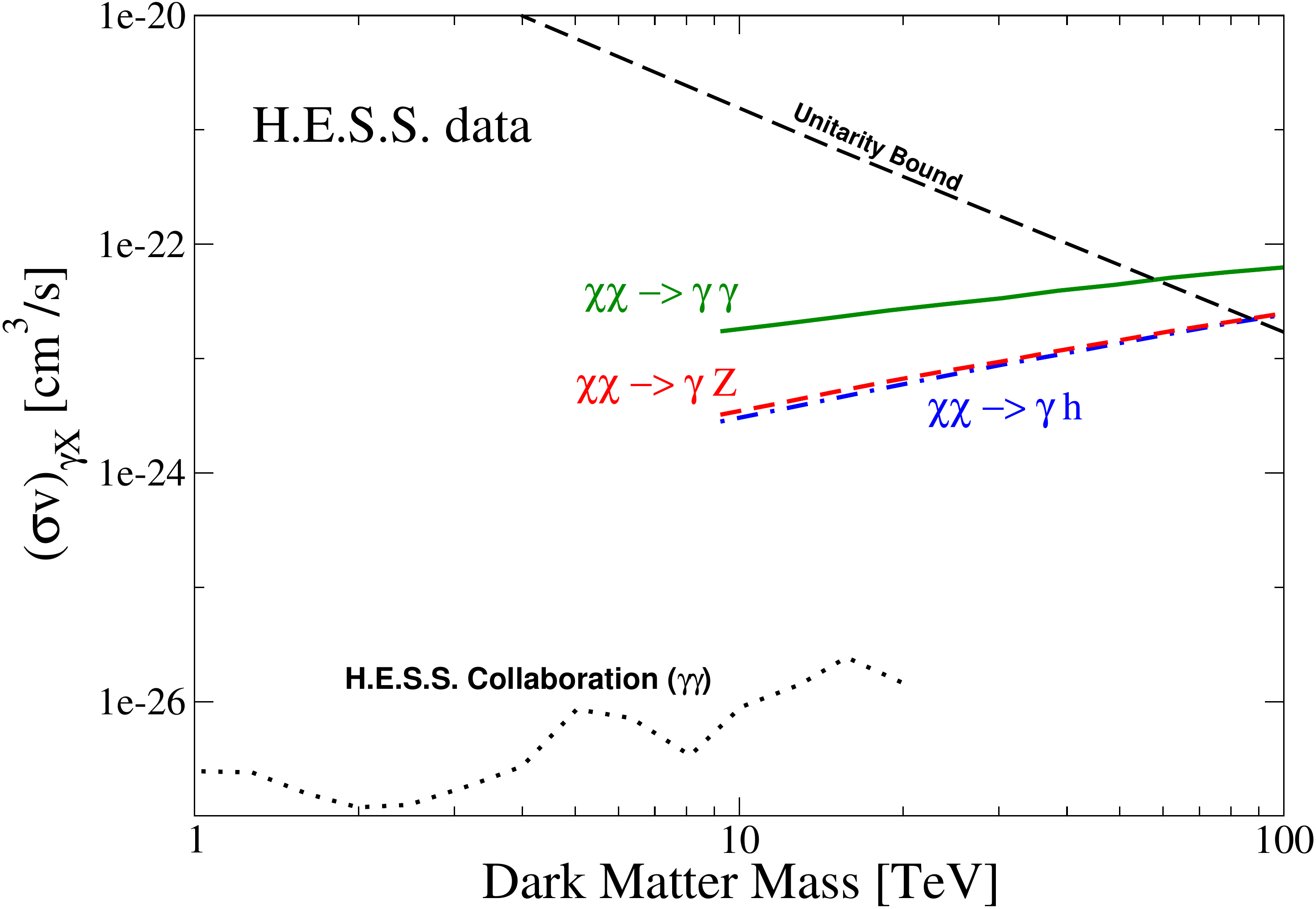}
\label{gammaline}
\caption{The Fermi-LAT and H.E.S.S. limits are represented by dotted curves in both figures, whereas the constraints for higher masses taking advantage of electroweak corrections are displayed in color.}
\end{figure}

\section*{Acknowledgments}

I thank the organizers of Moriond 2016 for the invitation to give the dark matter overview talk, and putting together great physicists in a lovely environment. I am grateful to Yann Mambrini, Stefano Profumo and Christoph Weniger for their collaboration in some of the results presented in this overview. I also thank Carlos Yaguna for his collaboration and proof-reading the manuscript. I am thankful to the organizers of the ``Dark Matter in the Milky Way'' program in Mainz for the mind-blowing workshop during which this review was partly written.




\end{document}